\documentclass[aps,prb,reprint,longbibliography, nofootinbib,superscriptaddress]{revtex4-2}

\usepackage{graphicx,amsmath,amssymb,color,amsfonts,bm,physics}
\usepackage{mathtools,lipsum}
\usepackage[dvipsnames]{xcolor}
\usepackage[misc]{ifsym}

\usepackage[colorlinks=true]{hyperref}  
\hypersetup{
    bookmarks=true,         
    unicode=false,          
    pdftoolbar=true,        
    pdfmenubar=true,        
    pdffitwindow=false,     
    pdfstartview={FitH},    
    pdftitle={},    
    pdfauthor={},     
    pdfsubject={},   
    pdfcreator={},   
    pdfproducer={}, 
    pdfkeywords={} {} {}, 
    pdfnewwindow=true,      
    colorlinks=true,       
    linkcolor=blue, 
    citecolor=blue,        
    filecolor=magenta,      
    urlcolor=blue,           
        breaklinks=true
} 

\newcommand{\be}{\begin{equation}}
\newcommand{\ee}{\end{equation}}
\newcommand{\bea}{\begin{eqnarray}}
\newcommand{\eea}{\end{eqnarray}}

\begin{document}

\title{Quantum chaos at finite temperature in local spin Hamiltonians }

\author{Christopher M. Langlett}
\affiliation{Department of Physics \& Astronomy, Texas A\&M University, College Station, TX 77843}

\author{Cheryne Jonay}
\affiliation{Faculty of Mathematics and Physics,
University of Ljubljana, 1000 Ljubljana, Slovenia}

\author{Vedika Khemani}
\affiliation{Department of Physics, Stanford University, Stanford, CA 94305}

\author{Joaquin F. Rodriguez-Nieva}
\affiliation{Department of Physics \& Astronomy, Texas A\&M University, College Station, TX 77843}


\begin{abstract}

Understanding the emergence of chaos in many-body quantum systems away from semi-classical limits, particularly in spatially local interacting spin
Hamiltonians, has been a long-standing problem. In these intrinsically quantum regimes, quantum chaos has been primarily understood through the correspondence between the eigensystem statistics of {\it midspectrum} eigenstates and the universal statistics described by random matrix theory (RMT). However, this correspondence no longer holds for finite-temperature eigenstates. Here we show that the statistical properties of finite-temperature eigenstates of quantum chaotic Hamiltonians can be accurately described  by pure random states constrained by a local charge, with the average charge density of the constrained random state ensemble playing the same role as the average energy density of the eigenstates. By properly normalizing the energy density using a single Hamiltonian-dependent parameter that quantifies the typical energy per degree of freedom, we find excellent agreement between the entanglement entropy statistics of eigenstates and that of constrained random states. Interestingly, in small pockets of Hamiltonian parameter phase space which we previously identified as `maximally chaotic' [PRX {\bf 14}, 031014~(2024)], we find excellent agreement not only at the level of the first moment, including O(1) corrections, but also at the level of statistical fluctuations. These results show that notions of maximal chaos---in terms of how much randomness eigenstates contain---can still be defined at finite temperature in physical Hamiltonian models away from semi-classical and large-$N$ limits.
\end{abstract}

\maketitle

\noindent{\bf Introduction.---}Describing how chaos emerges in quantum many-body systems has been a long-standing challenge~\cite{Deutsch1991Quantum, Srednicki1994Chaos, rigol2008thermalization, nandkishore2015many}.
In quantum systems exhibiting semiclassical limits, e.g. large-$N$ models~\cite{Sachdev1993Gapless,2017JHEP_sykchain2,roberts2018operator} or field theories~\cite{2014PRD_entanglementgrowth,Blake2021}, 
classical notions of chaos can be extended to quantum regimes by constructing quantum analogs of Lyapunov exponents~\cite{Avdoshkin2020Euclidean, Parker2019Universal,Maldacena_2016}.
These exponents quantify the growth of quantum state complexity under chaotic unitary evolution.
In such systems, the quantum Lyapunov exponent is well-defined due to the parametrically long temporal window in which complexity grows, and has been shown to reach an upper bound, which depends on temperature and fundamental constants, in systems exhibiting `maximally chaotic' behavior~\cite{Maldacena_2016}.

In generic quantum systems away from semiclassical limits, the correspondence with classical chaos no longer applies as the quantum Lyapunov exponent is no longer well-defined~\cite{2018PRL_quantumlyapunov}. 
In these regimes, quantum chaos is instead studied using the correspondence between the statistical properties of {\it midspectrum} eigenstates---those corresponding to infinite temperature in the thermodynamic limit---and the universal statistics described by random matrix theory~(RMT) ~\cite{Deutsch1991Quantum, Srednicki1994Chaos, rigol2008thermalization, nandkishore2015many}. This correspondence describes key `coarse-grained' features shared by {all} quantum chaotic systems away from integrable limits, independently of the microscopic details. These features include the level spacing statistics~\cite{Atas2013Distribution,atas2013joint,Oganesyan2007Localization} exhibiting a Wigner-Dyson distribution~\cite{gutzwiller2013chaos, Crossover2004Mills},
the spectral form factor becoming RMT distributed~\cite{Chan2018spectral,Friedman2019Spectral}, and mid-spectrum eigenstates displaying volume-law entanglement entropy~\cite{Renyi2019Lu,Vidmar2017Entanglement,Murthy2019Structure,2023PRE_averageEE,Bianchi2022Volume, langlett2024entanglement}.

\begin{figure}[t]
\includegraphics[width=\columnwidth]{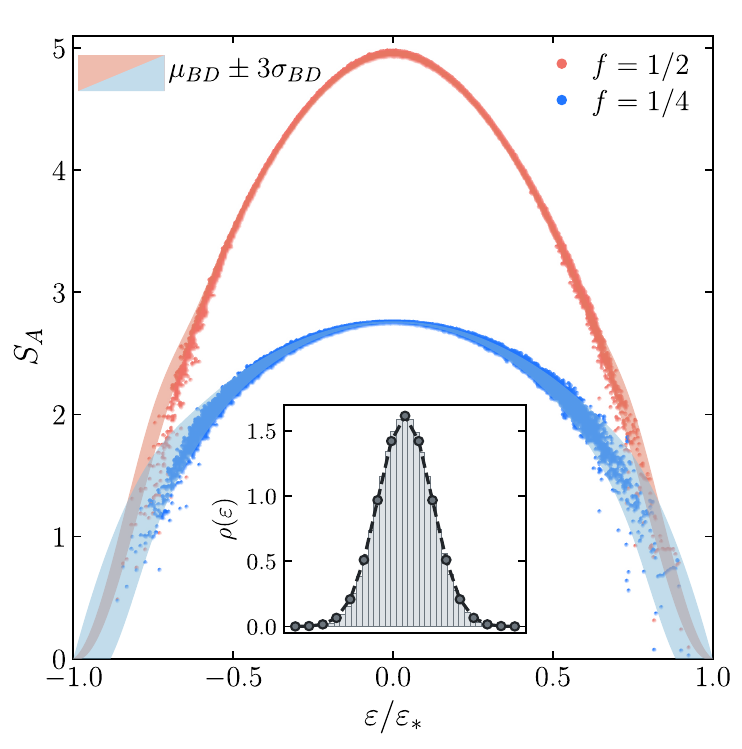}
\caption{The entanglement entropy~(EE) for each eigenstate of the quantum chaotic Hamiltonian in Eq.~(\ref{eq:MFIM}) for subsystem fractions $f=L_A/L=1/2$~(red) and $f=1/4$~(blue). The energy density $\varepsilon = E/L$ of each eigenstate is normalized with the average energy per qubit $\varepsilon_*$, which is obtained from the density of states~(see inset). The shaded regions indicate the mean $\pm$ three standard deviations of the exact EE distribution of pure random states with a U(1) conservation law~\cite{Bianchi2019Typical, Bianchi2022Volume} and represents the range within which approximately 99.7\% of the data points lie.
To generate the data points, we use the Mixed Field Ising Model (MFIM) with parameters $L=16$, $g=1.1$, $h=0.35$, and $\varepsilon_{*}=1.56$.
}
\label{fig:Fig1}
\end{figure}

The correspondence between quantum chaotic eigenstates and RMT ensembles, however, breaks down for eigenstates away from the middle of the spectrum (or, colloquially,  for `finite-temperature' eigenstates). In particular, finite-temperature eigenstates explore a smaller fraction of Hilbert space, thus they exhibit less entropy than midspectrum eigenstates and larger statistical fluctuations. Understanding these regimes is particularly important for the quantum simulation of low-temperature phases of matter, where emergent behaviors result from the interplay between thermal and quantum fluctuations\cite{2019Nature_RydbertKZ,2021Nature_RydbertAFM,2021Nature_Rydberg256,2024arxiv_rydbergcoarsening}. In these regimes, identifying the `appropriate' reference random state ensemble that should be used to diagnose and quantify chaos remains a challenge. 

Here we show that the statistical properties of finite-temperature eigenstates in strongly chaotic systems, specifically the entanglement entropy (EE) statistics of eigenstates in local spin Hamiltonians, can be accurately described by pure random states constrained by a local U(1) scalar charge,  whereby the charge density of the random state ensemble plays the same role as the energy density 
in the Hamiltonian system~[Fig.~\ref{fig:Fig1}]. These results generalize and extend the conclusions of our previous work~\cite{Rodriguez2014Quantifying}, which showed that locality combined with energy conservation imprints structure in the statistical properties of mid-spectrum (infinite temperature) eigenstates, to the finite temperature regime of quantum chaotic systems. By normalizing the energy density $\varepsilon = E/L$ of each eigenstate ($E$: energy relative to the middle of the spectrum) using a single Hamiltonian-dependent parameter $\varepsilon_*$ that quantifies the typical energy per qubit, we accurately describe the microcanonical distribution of eigenstate EE at finite temperature and for various subsystem sizes~(Fig.~\ref{fig:Fig1}).

Separately, we show that notions of `maximal chaos'---defined in terms of how much randomness eigenstates contain relative to the (appropriately) constrained random states---can also be extended to finite temperature eigenstates of spatially local spin-1/2 Hamiltonians. This is shown by studying the {\it fine-grained} statistics of the microcanonical eigenstate ensemble at finite temperature ~\cite{2019PRL_chalker,2019PRE_kurchan,2020PRE_beyondETH,2020PRA_higherorderETH,Garratt2021,2022PRL_dymarsky,2021PRE_eth_otocs,2022PRE_deviationsfromETH,2023PRX_cotleremergentdesign,2022PRL_exactdesigns,2023arxiv_fava,ghosh2024late}, particularly going beyond the widely-studied `volume-law' term. Close to the model parameters identified as maximally chaotic in Ref.~\cite{Rodriguez2014Quantifying}, we find that the entanglement patterns of finite-temperature eigenstates are accurately described by the constrained random states not only at the average level [including O(1) corrections] but also at the level of statistical fluctuations~(Fig.~\ref{fig:Fig1}). We provide extensive numerical evidence showing excellent convergence between ensembles in the thermodynamic limit, specifically at the level of the first two moments.  This work furnishes, for the first time, signatures of maximal chaos at finite temperature in a chaotic system that is spatially local and has a small local Hilbert space, i.e., having no semiclassical limit. 

The remainder of this work is structured as follows: we begin by reviewing the EE statistics of pure random states with a U(1) constraint, such as particle number conservation~(see Supplement for details). Next, we introduce the microscopic Hamiltonian, the mixed-field Ising model, where energy conservation combined with spatial locality gives rise to an effective local scalar charge: the energy density $\varepsilon $ (properly normalized) plays the same role as the charge density $n$ in the thermodynamic limit~\cite{Rodriguez2014Quantifying, Huang2021}. We then present extensive numerical results for the EE distribution across finite-energy eigenstates, considering both quarter- and half-subsystems, and discuss future outlook.

\noindent{\bf Constrained Entanglement Ensembles.---}We analyze the statistics of quantum states from the perspective of their entanglement entropy $S_{A} = -\text{Tr}[\rho_{A}\log(\rho_{A})]$, where $\rho_A$ is the reduced density matrix for a subsystem $A$ with $L_A$ qubits.
For a system with a local scalar charge and a local Hilbert space dimension $d=2$, it is convenient to think of $0 \leq N \leq L$ as an integer charge number where each site accommodates a maximum of a single charge. In this case, the Hilbert space ${\mathcal{H}(N)}$ for a fixed charge $N$ decomposes as a direct sum of tensor products, ${\cal H}(N) = \bigoplus_{N_A} {\cal H}_A(N_A)\otimes{\cal H}_B(N-N_A)$, where $N_A$ is an integer within the range ${\rm max}(0, N-L_B) \le N_A \le {\rm min}(N, L_A)$. A random state within a fixed charge sector $\ket{\Phi_N} \in {\cal H}(N)$ is expressed as a superposition of orthonormal basis states, $|\Phi_N\rangle = \sum_{N_A,\alpha,\beta}\phi_{\alpha,\beta}^{(N_A)}|N_A,\alpha\rangle\otimes|N-N_A,\beta\rangle$, with $\phi^{(N_A)}_{\alpha,\beta}$ uncorrelated random numbers up to normalization. The index $\alpha$~($\beta$) labels the basis states in subsystem $A$~($B$) with a total charge $N_A$~($N-N_A$). The reduced density matrix of the constrained random states over subsystem $A$ is block diagonal $\rho_{A}= {\rm Tr}_B[|\Phi_N\rangle\langle\Phi_N|]= \sum_{N_A} p_{N_A} \rho_{A|N_A}$, where $\rho_{A|N}$ denotes the block with $N_A$ particles, and the factors $p_{N_A}\ge 0$ are the (classical) probability distribution of finding $N_A$ particles in $A$. The entanglement entropy has the form $S(\rho_{A}) = \sum_{N_A} p_{N_A} S(\rho_{A|N_A})- p_{N_A} \log p_{N_A}$, where the second term on the RHS is the Shannon entropy of the number distribution $p_{N_A}$, which captures particle number correlations between the two subsystems, and the first term captures quantum correlations between configurations with a fixed particle number. 

The first two moments of the EE distribution for the ensemble of pure random states with constrains were first computed exactly by Bianchi and Dona in Ref.~\cite{Bianchi2019Typical, Bianchi2022Volume}, therefore generalizing the analytical results for Haar random states that was first conjectured by Page~\cite{1993PRL_Page} and later proven by others~\cite{2016PRE_entanglementdispersion,2017PRE_entanglementvariance_proof} (we note that a previous work derived the asymptotic behavior of the first moment of EE for fixed charge number~\cite{Vidmar2017Entanglement}). Similarly to the notation we used in Ref.~\cite{Rodriguez2014Quantifying}, we refer to the constrained distribution the Bianchi-Dona (BD) distribution. 
For clarity, we only discuss the asymptotic results for the first two moments in the main text and detail the exact expressions (which are used in the figures) in the Supplement.
The average EE in the asymptotic limit is
\begin{align}
\mu_{\text{BD}}(n,f) = & -\left[n\log n+(1-n)\log(1-n)\right]fL \nonumber \\
& -\sqrt{\frac{n(1-n)}{2\pi}}\left| \log\left(\frac{1-n}{n}\right) \right|\delta_{f,1/2}\sqrt{L} \label{eq:muM} \\
& + \frac{f+\log(1-f)}{2}-\frac{1}{2}\delta_{f,1/2}\delta_{n,1/2}\nonumber,
\end{align}
where $n = N/L$ is the charge density and $f=L_A/L$ is the subsystem fraction.
The first and third terms on the RHS arise from the mean-field behavior of the EE defined as $\mu_{\rm MF} = -{\rm Tr}[\bar{\rho}_A\log\bar{\rho}_A]$, with $\bar{\rho}_A$ the average density matrix within a given charge sector $N$.
The second and fourth terms arise from fluctuations within particle subsectors and are nonzero only when considering half-systems $f=1/2$. Further, the second term is non-zero when the system is away from half-filling, i.e., $n \neq 1/2$.

The analytic expression for the EE fluctuations within a sector $N$ is cumbersome~(see Supplement), but its asymptotic behavior near $n=1/2$ is simple:
\be
\sigma_{\text{BD}}(n,f) \approx \frac{{L}^{1/4}}{D_n^{1/2}}G_{f}(\sqrt{L} \delta n), \quad \delta n = n-1/2.
\label{eq:stdM}
\ee
Here $G_f(x)$ is a smooth $f$-dependent function that is O(1)~(see the Supplement), and $D_n=e^{-[n\log n + (1-n)\log(1-n)]L}$. 
The prefactor $(L^{1/4}/D_n^{1/2})$ in Eq.~(\ref{eq:stdM}) comes from using Stirling's approximation to quantify the Hilbert space dimension at a finite charge density, while the $\sqrt{L}$ in the argument of $G_f(x)$ ensures scale invariance of $G_f$ by accounting for the reduction in charge fluctuations with increasing system size, characterized by $\langle n^2\rangle = 1/L$.

We now return to spatially local Hamiltonians. We first note that caution is required when attempting to compare the EE statistics of Hamiltonian eigenstates with that of pure random states constrained with a U(1) scalar charge, as energy eigenstates factor as $|\Phi_E\rangle \approx \sum_{E_A}\sum_{\alpha\beta}\phi_{\alpha,\beta}^{(E_A)}|E_A,\alpha\rangle\otimes|E-E_A,\beta\rangle$ approximately but not exactly (here $|E_{A},\alpha\rangle$ and $|E_B,\beta\rangle$ are the Hamiltonian eigenstates within subsystems $A$ and $B$, respectively). 
This is because of Hamiltonian terms that couple subsystems $A$ with $B$, which effectively `smear' the charge within the subregions. For mid-spectrum eigenstates, the smearing of the local charge has two competing effects. On the one hand, it tends to increase EE because a larger fraction of Hilbert space is explored; on the other hand, it tends to decrease EE because states with smaller charge density have lower EE. 
In Ref.~\cite{Rodriguez2014Quantifying}, we argued that both effects compensate in such a way that the numerical results for the EE distribution in strongly chaotic Hamiltonians do not display significant deviations from Eq.~(\ref{eq:muM}) within the scale defined by $\sigma_{\rm BD}$ in Eq.~(\ref{eq:stdM}).

In contrast, away from the middle of the spectrum, the smearing of the charge tends to increase the EE, as the two effects discussed above now act towards increasing the exploration of Hilbert space. 
In the supplement we show that the smearing of a U(1) charge for $N<L/2$ does not significantly increase the average EE on the scale of the standard deviations, so long as $\delta N \lesssim O(1)$. As such, Eqs.~(\ref{eq:muM})-(\ref{eq:stdM}) are also expected to accurately describe finite-energy eigenstates. This observation agrees with the numerical behavior observed below for Hamiltonian systems.

\noindent{\bf Quantum Chaotic Model.---}We consider the one-dimensional mixed-field Ising model~(MFIM), which has been widely used as a canonical model of quantum chaos~\cite{2011PRL_banulscirachastings,Ballistic2013Kim,Thermalization2015Zhang,roberts2015localized}: 
\begin{equation}
    \label{eq:MFIM}
    H = \sum_{i=1}^L \left(Z_{i}Z_{i+1} + g X_{i} + h Z_{i}\right).
\end{equation}
Here, $\{X_i, Y_i, Z_i \}$ are the Pauli operators acting on each site $i$ of a chain of $L$ qubits, $g$ is the transverse field, and $h$ is the longitudinal field. 
The Hamiltonian Eq.~\eqref{eq:MFIM} has additional point symmetries, which we explicitly break.
In particular, to break translation invariance, we use open boundary conditions and we include the boundary terms $h_1=Z_1 /4$ and $h_N = -Z_N /4$ at the edges of the chain to break inversion symmetry. As a result, the only remaining symmetry in Eq.~\eqref{eq:MFIM} is energy conservation, combined with spatial locality. 

The MFIM exhibits various regimes in parameter space. When the longitudinal field is zero, $h=0$, the model can be mapped to free-fermions by using a Jordan-Wigner transformation~\cite{Two1964Lieb}.
The MFIM also hosts two classical limits: (i) when $g=0$ the model becomes the classical Ising model~(diagonal in the $Z$ basis), 
and (ii) when $g\gg 1$ the model becomes the classical paramagnet~(diagonal in $X$ basis). 
Here we are primarily interested in regimes away from all these integrable limits, particularly in the proximity of
$(g, h) = (1.1, 0.35)$ which we identified as the most chaotic parameters in Ref.~\cite{Rodriguez2014Quantifying}, and is also close to the parameters widely used in the literature when studying quantum chaos~\cite{2011PRL_banulscirachastings}.

\noindent{\bf Single parameter correspondence between ensembles.---}To make a one-to-one comparison between the eigenstate statistics of local Hamiltonians and pure random states constrained with a U(1) scalar charge, we first define an energy scale $\varepsilon_*$ that quantifies the average energy per qubit. For this purpose, we first note that the density of states in local Hamiltonian systems and in systems with a U(1) scalar charge become Gaussian distributed in the thermodynamic limit. In a system with total charge $N = \frac{1}{2}\sum_{i=1}^L (1-Z_i)$, the average charge is $\mu_N = {\rm Tr}[N] = L/2$, and the charge variance is $\sigma_N^2 = \langle N^2 \rangle - \langle N\rangle^2= L/4$. Similarly, the Hamiltonian (\ref{eq:MFIM}) has average energy $\langle H \rangle = {\rm Tr}[H] = 0$ because $H$ is written as a sum of Pauli matrices, and the energy variance is $\delta H^2 = {\langle H^2 \rangle} = L(J^2+g^2+h^2)=L\varepsilon_*^2$ (here we neglected the contribution of the boundary terms, but we include these corrections in the values of $\varepsilon_*$ used in the figures). We emphasize that the energy scale $\varepsilon_*$ is a `classical' energy scale in the sense that it has no information about the entanglement structure of eigenstates. In what follows, we express all our numerical data in terms of the energy density of the n-th eigenstate, $\varepsilon_n = E_n/L $ normalized by $\varepsilon_*$, where the variable $\varepsilon_n/\varepsilon_*$ plays the same role as the charge density $0 \le n = N/L \le 1$ through the correspondence $-1 \le \varepsilon/\varepsilon_* \equiv 2n-1 \le 1$.

\begin{figure}[t]
\includegraphics[width=\columnwidth]{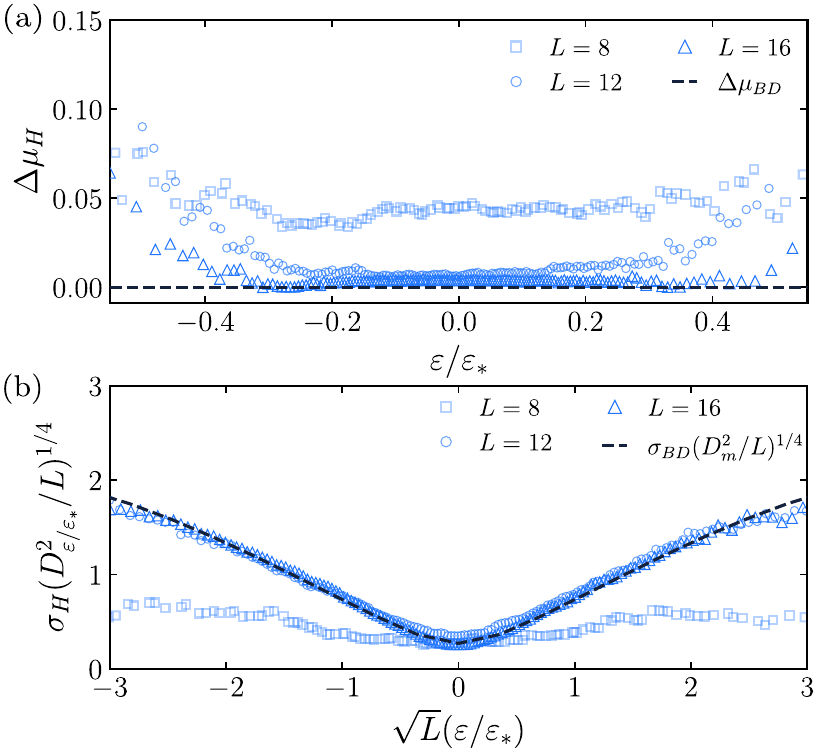}
\caption{Finite-size scaling of the microcanonical EE distribution for $f=1/4$ and $L = 8, 12, 16$. The data is shown in terms of the (a) first and (b) second moment of the microcanonical EE distribution computed for energy bins of size $\Delta \varepsilon = 0.005\varepsilon_*$. For clarity, in panel (a) we show the data relative to the mean-field EE, $\Delta\mu_H = \mu_H - \mu_{\rm MF}$, see discussion following Eq.~(\ref{eq:muM}). In panel (b), we rescale the $x$ and $y$ axes such that all data points collapse into a single universal function $G_{1/4}(x)$ described by Eq.~(\ref{eq:stdM}). 
The dashed lines in (a) and (b) are the exact asymptotic values for the constrained random state ensemble as $L\rightarrow\infty$~\cite{Bianchi2019Typical}. The same model parameters as in Fig.~\ref{fig:Fig1} are used.
}
\label{fig:Fig2}
\end{figure}

\noindent{\bf Numerical Results for Quarter Subsystem.---}We analyze the raw data in Fig.~\ref{fig:Fig1} moment by moment for $f=1/4$ and as a function of system size $L = 8, 12, 16$, see Fig.~\ref{fig:Fig2}. For this purpose, we use a small energy window of size $\Delta \varepsilon = 0.005\varepsilon_*$ to compute the microcanonical distribution of eigenstate EE as a function of $\varepsilon$. For the largest system size $L=16$, the energy window captures approximately 500 eigenstates in the middle of the spectrum, and this number decreases as we move away from zero energy, but we always keep a statistically significant number of eigenstates in each window across the energy range shown in Fig.~\ref{fig:Fig2}. Figure~\ref{fig:Fig2}(a) shows the microcanonical average
of the eigenstate EE across windows centered at different energies. For clarity, we show $\Delta \mu_{H} = \mu_{H} - \mu_{\rm MF}$ where we have subtracted the volume-law term and the O(1) correction,
$ \mu_{\rm MF} = -L_A[n\log n + (1-n)\log(1-n)] - \frac{f + \log(1-f)}{2} $, from the average EE, both of which stem from the mean-field EE~\cite{Vidmar2017Entanglement}. The expectation for $f<1/2$ is that $\Delta \mu_H \rightarrow 0$ in the thermodynamic limit (dashed line), which agrees remarkably well with the numerical behavior of the eigenstate data.

Figure~\ref{fig:Fig2}(b) shows the second moment of the EE distribution, $\sigma_{H}$, as a function of $\varepsilon$. To show the excellent collapse of the data points as a function of system size, we rescale the value of $\sigma_{H}$ ($y$ axis) by dividing with the prefactor $(L/D_{\varepsilon/\varepsilon_*}^2)^{1/4}$, see Eq.~(\ref{eq:stdM}). In addition, we rescale the $x$ axis with the prefactor $\sqrt{L}$ to account for the decrease of energy variance as the system size increases. The dashed line indicates the asymptotic behavior of the (properly rescaled) EE fluctuations for the U(1) conserving systems (see Supplement). We find excellent collapse of the data points for all system sizes, especially excellent agreement at the level of the asymptotic function $G_{1/4}(x)$ in Eq.~(\ref{eq:stdM}). 

\begin{figure}
\includegraphics[width=\columnwidth]{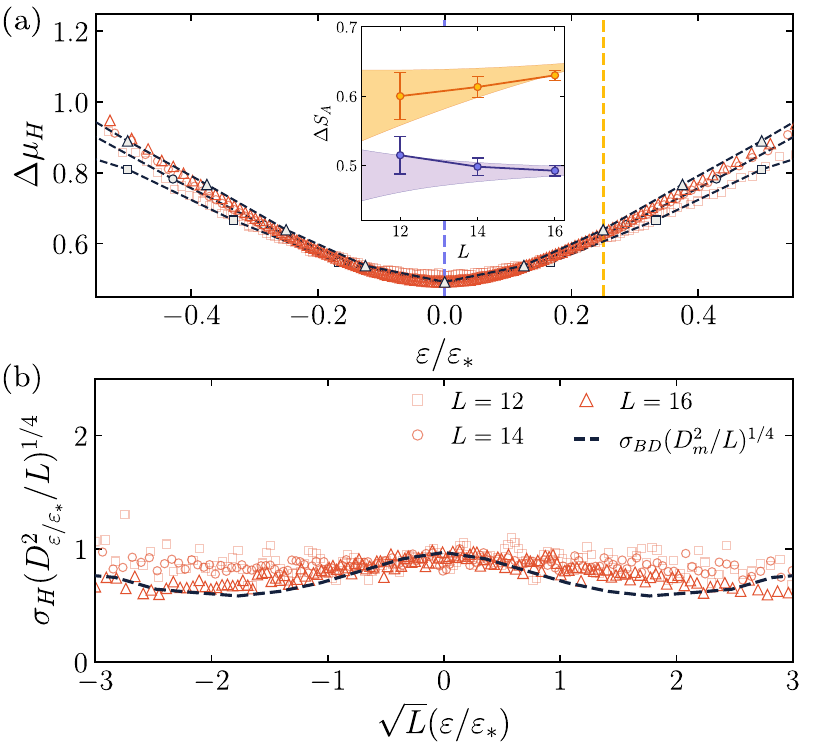}
\caption{Finite-size scaling of the microcanonical EE distribution for $f=1/2$ and $L=12,14,16$. As in Fig.~\ref{fig:Fig2}, the data is shown in terms of the (a) first and (b) second moment of the EE distribution computed for energy bins of size $\Delta \varepsilon = 0.005\varepsilon_*$. In panel (a) we show the data relative to the mean-field EE value, and in panel (b) we rescale the $x$ and $y$ axes such that all data points collapse into a single universal function $G_{1/2}(x)$, see Eq.~(\ref{eq:stdM}). We note that the scaling of $\Delta \mu_H$ with  system size at zero energy is different from that at finite energies, see Eq.(\ref{eq:muM}). For this reason, in the main panel we plot the exact values of $\Delta \mu_H$ obtained for pure random states with a U(1) constraint 
at {\it finite} $L$~(dashed lines). The inset of panel (a) shows the system-size dependence of $\Delta \mu_H$ for fixed $\varepsilon/\varepsilon_* = 0$ and $\varepsilon/\varepsilon_* = 0.25$. The same model parameters as in Fig.~\ref{fig:Fig1} are used.
}
\label{fig:Fig3}
\end{figure}

\noindent{\bf Numerical Results for Half Subsystem.---}We now analyze the raw data in Fig.~\ref{fig:Fig1} moment by moment for $f=1/2$ and as a function of system size $L = 12, 14, 16$, see Fig.~\ref{fig:Fig3}. We use the same partitioning into small energy windows as that used in Fig.~\ref{fig:Fig2}. Figure \ref{fig:Fig3}(a) shows the microcanonical average of the eigenstate EE, where we also removed the mean-field term as in Fig.~\ref{fig:Fig2}(a). Unlike Fig.~\ref{fig:Fig2}(a), however, it is more challenging to show the approach of the numerical data towards the expected EE distribution because different energy densities exhibit different scaling behavior: (i) when $\varepsilon/\varepsilon_* = 0$, then $\Delta \mu_H = 1/2$ and, (ii) when $\varepsilon/\varepsilon_*$ is finite, then $\Delta \mu_H \propto \sqrt{L}$. For this reason, we plot the exact value of $\Delta \mu_{\text{BD}}$ for each system size (see Supplement). We find that the numerical data converges remarkably well towards the constrained random state ensemble for the numerically accessible system sizes. 

To more quantitatively show the convergence of $\Delta \mu_H$ towards the theoretical prediction, we also plot the system size dependence of the average EE at a fixed value of $\varepsilon/\varepsilon_*$ in the inset of Fig.~\ref{fig:Fig3}(a). We specifically consider $\varepsilon/\varepsilon_* = 0 $ (mid-spectrum eigenstates) and $\varepsilon/\varepsilon_* = 0.25 $ (finite temperature eigenstates) as a function of system size. The error bar indicates the variance of the EE distribution within a given energy window, whereas the shaded areas indicate the region limited by $\mu_{\text{BD}}\pm\sigma_{\text{BD}}$ at the corresponding charge density. We find very good agreement between the numerical values of EE and the theoretical prediction for the available system sizes, not only at the level of averages but also the fluctuations. 

Finally, Fig.~\ref{fig:Fig3}(b) shows the second moment of the microcanonical EE distribution for $f=1/2$. We rescale the data in exactly the same way as done in Fig.~\ref{fig:Fig2}(b), and the dashed lines indicate the asymptotic behavior of the EE fluctuations of the constrained random state ensemble. Similarly to the $f=1/4$ case, we observe an excellent collapse of the data points towards the expected scaling function $G_{1/2}(x)$ in Eq.~(\ref{eq:stdM}).

\noindent{\bf Discussion.---}Our results furnish a diagnostic of quantum chaos at finite temperature in physical Hamiltonian models away from semi-classical and large $N$ limits, and reveal stricking signatures of maximal chaos in these intrinsically quantum regimes.  The essential step in our analysis comes from studying the fine-grained features of Hamiltonian eigenstates, which uncovers a richer entanglement structure than what is revealed from the average `volume-law' behavior of the entanglement entropy. Looking ahead, it would be interesting to investigate how these signatures of maximal chaos extend to zero energies, particularly at the transition between volume-law and area-law entanglement behaviors~\cite{Eigenstate2021Miao,Scaling2021Barthel}, and through different entanglement observables beyond the von Neumann entropy\cite{2020PRB_Grover_tripartite}. 
Additionally, exploring the dynamical signatures of maximal chaos at finite energies, as well as probing chaos at finite charge densities in systems with more complex symmetries, such as non-Abelian symmetries~\cite{2023PRL_su2chaitanya,2023PRB_su2ee,Patil2023Average,bianchi2024non}, or in the presence of symmetry breaking~\cite{russotto2024non}, are interesting avenues for future research.

JFRN acknowledges the hospitality of the Aspen Center for Physics, which is supported by National Science Foundation grant PHY-2210452, and a grant from the Alfred P. Sloan Foundation (G-2024-22395). This work was also supported by the US Department of Energy, Office of Science, Basic Energy Sciences, under Early Career Award Nos. DE-SC0021111 (C.J. and V.K.).  V.K. also acknowledges support from the Alfred P. Sloan Foundation through a Sloan Research Fellowship and the Packard Foundation through a Packard Fellowship in Science and Engineering. The numerical simulations in this work were conducted with the advanced computing resources provided by Texas A\&M High Performance Research Computing. 

%

\clearpage

\renewcommand{\thefigure}{S\arabic{figure}}
\renewcommand{\theequation}{S\arabic{equation}}
\renewcommand{\thesection}{S\arabic{section}}
\setcounter{page}{1}
\setcounter{equation}{0}
\setcounter{figure}{0}
\setcounter{section}{0}

\begin{widetext}

\begin{center}
{\large\underline{\bf SUPPLEMENTARY MATERIAL} \\ {\bf Quantum chaos at finite temperature in local spin Hamiltonians}

\author{}}

\vspace{4mm}

Christopher M. Langlett,$^1$ Cheryne Jonay,$^2$ Vedika Khemani,$^3$ and Joaquin F. Rodriguez-Nieva$^1$

{\small\it $^1$Department of Physics \& Astronomy, Texas A\&M University, College Station, TX 77843}

{\small\it $^2$Faculty of Mathematics and Physics, University of Ljubljana, 1000 Ljubljana, Slovenia}

{\small\it $^3$Department of Physics, Stanford University, Stanford, CA 94305}

\end{center}

\end{widetext}

\begin{figure}
\includegraphics[width=\columnwidth]{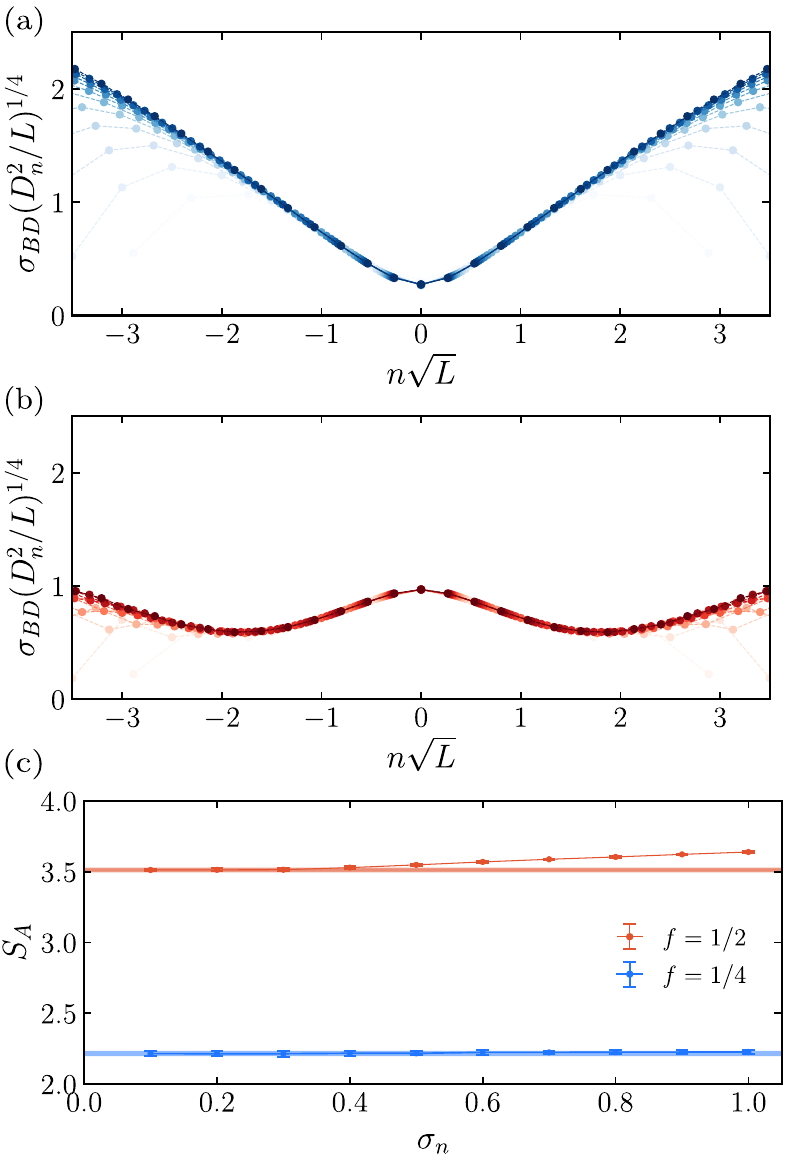}
\caption{Finite-size scaling of the EE fluctuations, Eq.~(\ref{eq:BDsdevexact}), shown for subsystem sizes (a) $f=1/4$ and (b) $f=1/2$. Here $\delta n$ denotes $\delta n = n -1/2$. In both panels, we rescale the $x$ and $y$ axes according to Eq.(\ref{eq:stdM}) such that all data points collapse onto a single universal function $G_{1/4}(x)$ and $G_{1/2}(x)$. The system sizes used in (a) and (b) are $L=[12, 16, \ldots, 60]$ in steps of $\Delta L = 4$. 
(c) Distribution of EE as a function of particle number fluctuations $\sigma_N$ for $n = 1/4$, and system fractions $f=1/4$ and $f=1/2$. The error bars indicate one standard deviation within the mean EE value. The shaded regions indicate the range $\mu_{\rm BD}\pm\sigma_{\rm BD}$ of the microcanonical distribution for a fixed magnetic charge $n=1/4$, and fractions $f=1/4$ (blue) and $f=1/2$ (red).
}
\label{fig:SMFig1}
\end{figure}

This Supplementary Material discusses the analytical details of the entanglement distribution for pure random states constrained by a U(1) charge for both the first and second moment. We then numerically perform a finite-size analysis on the exact results illustrating the collapse to a universal function for both $f=1/4$ and $f=1/2$. The final section numerically studies the EE distribution of finite energy eigenstates of the MFIM away from the maximally chaotic point.

\section{Constrained random state ensembles with a U(1) Charge}
\label{sec:BianchiDist}

The statistical properties of EE for pure random states constrained by a U(1) symmetry was first derived exactly by Bianchi and Dona. An excellent review is presented in Ref.~\cite{Bianchi2022Volume}. Here we recount the steps of deriving the first and second moment of the entanglement distribution that is most relevant to our work.

To begin, consider a chain of $L$ sites with a total number of $N$ particles, with $0 \le N \le L$, and each site able to accommodate a maximum of one particle. When the system is partitioned into two subsystems of sizes $L_A$ and $L_B$, the Hilbert space factors into
\begin{equation}
    \mathcal{H}(N) = \bigoplus_{N_A={\rm max}(0,N-L_B)}^{{\rm min}(N,L_A)} \mathcal{H}_{A}(N_A) \otimes \mathcal{H}_{B}(N-N_A).
\end{equation}
The total Hilbert space dimension of each $N$ particle sector is $d_N = \binom{L}{N}$, and the dimensions of each subsystem are $d_{N_A} = \binom{L_A}{N_A}$ and $d_{N_B}  = \binom{L-L_A}{N - N_A}$.

To determine the EE of a pure random state constrained to a symmetry sector, we begin with the random states, $|\Psi_N\rangle \in {\cal H}(N)$, defined as
\be
|\Psi_N\rangle = \sum_{N_A}\sum_{\alpha,\beta}\psi_{\alpha,\beta}^{(N_A)}|N_A,\alpha\rangle\otimes|N-N_A,\beta\rangle,
\ee
with coefficients $\psi_{\alpha,\beta}^{(N_A)}$ which are independently and identically distributed complex~(GUE) Gaussian variables.
The reduced density matrix of subsystem $A$ is block diagonal, $\rho_{A|N}= \sum_{N_A} p_{N_A} \rho_{A|N_A}$, and the factors $p_{N_A}\ge 0$ are the (classical) probabilities of finding $N_A$ particles in $A$, $p_{N_A} = \frac{d_{N_A}d_{N_B}}{d_N}$.

The EE can be written as 
\be
S(\rho_{A|N}) = \sum_{N_A} p_{N_A} S(\rho_{A|N_A})- p_{N_A} \log p_{N_A},
\ee
where the second term on the RHS is the Shannon entropy of the number distribution $p_{N_A}$, which captures particle number correlations between the two halves, while the first term is the Page entropy for the block with $N_A$ particles in subsystem $A$. 

The first moment of EE, $\mu_{\rm BD} = \langle S_A \rangle_N$,  is given by
\begin{align}
\label{eq:BDAverageExact}
    \mu_{\rm BD} =& \sum_{N_A} p_{N_A} \phi_{N_A}, \\ 
    \phi_{N_A} =& \Psi(d_N +1) - \Psi(\max(d_{N_A},d_{N_B})+1)\nonumber \\ &-\min\left(\frac{d_{N_A}-1}{2d_{N_B}},\frac{d_{N_B}-1}{2d_{N_A}}\right).
    \label{seq:phifunction}
\end{align}
In other words, the mean EE of states in a given U(1) symmetry sector is the Page entropy for all random blocks $\rho_{A|N_A}$ averaged with $p_{N_A}$. In the above $\Psi(x)$ is the digamma function which is the logarithmic derivative of the Gamma function.

The variance of the entanglement distribution for complex random states restricted to a symmetry sector $N$ is given by:
\begin{align}
\label{eq:BDsdevexact}
\sigma_{\rm BD}^2= \frac{1}{d_N +1}\bigg[\sum_{N_A}p_{N_A}\left(\phi_{N_A}^{2}+\chi_{N_A} \right) - \langle S_A \rangle_{N}^{2}\bigg],
\end{align}
where $p_{N_A}$, and $\phi_N$ are defined in the previous equations, and $\chi_N$ for $d_A \leq d_B$ takes the form,
\begin{widetext}
\be
\chi_{N_A} = 
\begin{cases}
     (d_{N_A}+d_{N_B})\Psi^{\prime}(d_{N_B}+1)-(d_N+1)\Psi^{\prime}(d_N+1) - \frac{(d_{N_A}-1)(d_{N_A}+2d_{N_B}-1)}{4d^{2}_{N_B}},\quad\text{if $d_{N_A}\le d_{N_B}$}\\
    (d_{N_A}+d_{N_B})\Psi^{\prime}(d_{N_A}+1)-(d_N+1)\Psi^{\prime}(d_N+1) - \frac{(d_{N_B}-1)(d_{N_B}+2d_{N_A}-1)}{4d^{2}_{N_A}},\quad\text{if $d_{N_A}>d_{N_B}$}.
\end{cases}
\label{eq:BDsdevexact2}
\ee
\end{widetext}
In the thermodynamic limit $L\rightarrow \infty$ with subsystem fraction $f=L_{A}/L$ and particle density $n=N/L$ fixed the above exact expressions can be evaluated with saddle point methods to derive the asymptotic forms in the main text.

\section{Scaling law behavior of the EE fluctuations}

To find the scaling law behavior of $\sigma_{\rm BD}$ for large systems, we first note that the term in brackets in Eq.(\ref{eq:BDsdevexact}) is O(1), thus the system size dependence of $\sigma_{\rm BD}^2$ is primarily dominated by the term $d_N$ in the denominator. Using Stirlings approximation, the denominator of Eq.(\ref{eq:BDsdevexact}) scales as $d_N \propto {e^{-[n\log n + (1-n)\log(1-n)]L}}/{\sqrt{L}}$. Upon rescaling, we note that the O(1) function still has an $L$ depedence in the argument, as the fluctuations of magnetization scales as $\langle n^2 \rangle \sim 1/L$ in the thermodynamic limit. 

This motivates the scaling law proposed in Eq.(\ref{eq:stdM}), which is valid in the asymptotic limit. To show that the scaling behavior works, we show that the exact datapoints obained from Eq.(\ref{eq:BDsdevexact}) collapse remarkably well using the scaling in Eq.(\ref{eq:stdM}).

\begin{figure}
\includegraphics[width=\columnwidth]{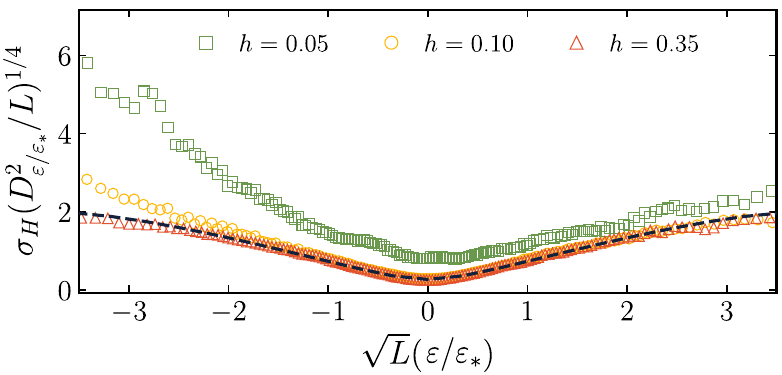}
\caption{Rescaled fluctuations of EE as a function of the $\varepsilon$ and the transverse field $h$, shown for $f=1/4$. We rescale the $x$ and $y$ axes according to Eq.(\ref{eq:stdM}) such that all data points collapse onto a single universal function $G_{1/4}(x)$. 
The parameters used are $L=16$, $g=1.1$, $\varepsilon_{*}=1.56$, and $\Delta \varepsilon=0.005\varepsilon_{*}$.
}
\label{fig:SMFig2}
\end{figure}

\section{Effect of smearing a local charge}

A key assumption of our work is that energy eigenstates can be approximated as random superpositions of states with a fixed number of local `energy qubits,' akin to the behavior observed in systems with a local magnetic charge $Z = \sum_i Z_i$. This assumption holds exactly for the terms $H_0 = \sum_i gX_i + h Z_i$ in Eq.(\ref{eq:MFIM}), but becomes approximate upon including the Ising interaction $H_1 = \sum_i JZ_iZ_{i+1}$. The Ising term introduces energy fluctuations within subsystems, characterized by $\delta E_A \sim O(1)$. 

For finite-energy eigenstates, one might expect that `smearing' the local charge across subsystems increases the entanglement entropy (EE) by allowing access to a larger portion of the Hilbert space. However, our numerical results for Hamiltonian systems indicate that this increase in EE due to smearing is small relative to the scale of the EE fluctuations, even when $J = 1$.

To support this observation, we quantitatively analyze a simpler scenario: adding fluctuations of total particle number. Instead of considering pure random states within a fixed sector $N$, we instead consider pure random states where each $N$-sector is assigned a probability $p_N$, with $p_N$ following a Gaussian distribution with standard deviation $\sigma_N$. This smearing of the U(1) charge serves as an analog to introducing the $J$ term in the Hamiltonian (\ref{eq:MFIM}) while preserving the global U(1) charge.

Figure \ref{fig:SMFig2} illustrates the behavior of the entanglement entropy (EE) as a function of $\sigma_N$ for a finite energy density $n = 1/4$ at $f = 1/4$ and $f = 1/2$. The error bars represent the range corresponding to one standard deviation around the mean EE value. For $f = 1/4$, we observe that the average EE does not increase significantly compared to its standard deviation, even for relatively large values of $\sigma_N \sim 1$. Similarly, for $f = 1/2$, the EE remains within the standard deviation of the BD distribution as long as $\sigma_N \lesssim 0.5$.
This insensitivity of the mean EE to the smearing of the magnetic charge for relatively large values of $\sigma_N \sim O(1)$ justifies why the distribution of pure random states with local constraints accurately captures the finite-energy eigenstates described in the main text.

\section{Deviations from maximal chaos}

All the results discussed in the main text are for the MFIM at the most chaotic point $(g,h) = (1.1,0.35)$. We now present numerical results for the finite-energy eigenstate statistics away from this points. In particular, we show that that deviations from the most chaotic parameters lead to non-universal statistics that are sensitive to model parameters. We will focus on the second moment of the EE distribution as a function of energy for $f=1/4$. Figure~\ref{fig:SMFig2} extends the results of Fig.~\ref{fig:Fig2}(b) to the Hamiltonian data with $h = 0.1$ and $h = 0.05$. We see that, as we depart from teh most chaotic parameter, larger deviations from the constrained random states emerge, even when the system is within the quantum chaotic regime.

\end{document}